\documentclass[aps,prb,twocolumn,groupedaddress]{revtex4}
\usepackage{graphicx}
\newcommand{\LiTiO}{LiTi$_2$O$_4$}
\newcommand{\LiVO}{LiV$_2$O$_4$}
\newcommand{\LixTiO}{Li$_{1+x}$Ti$_{2-x}$O$_4$}
\newcommand{\LiAlTiO}{LiAl$_y$Ti$_{2-y}$O$_4$}

\begin{document}

\title{Examining the metal-to-insulator transitions \\
in \LixTiO~and \LiAlTiO \\ with a Quantum Site Percolation model}


\author{}
\author{F. Fazileh}
\author{R. J. Gooding}
\affiliation{Department of Physics, Queen's University, Kingston ON K7L 3N6 
CANADA}
\author{D. C. Johnston}
\affiliation{Ames Laboratory and Department of Physics and Astronomy,
Iowa State University, Ames, Iowa 50011 USA}


\date{\today}

\begin{abstract}
We have studied the composition-induced metal-to-insulator transitions 
(MIT) of cation substituted Lithium
Titanate, in the forms \LixTiO~and \LiAlTiO, utilising a
quantum site percolation model, and we argue that such a model provides
a very reliable representation of the noninteracting electrons in this material
{\em if} strong correlations are ignored.
We then determine whether or not such a model of $3d^1$ electrons
moving on the Ti (corner-sharing tetrahedral) sublattice describes the observed
MITs, with the critical concentration defined by the matching of the mobility 
edge and the chemical potential. Our analysis leads to quantitative predictions 
that are in disagreement with those measured experimentally. For example, 
experimentally for the \LiAlTiO~compound an Al concentration of
$y_c \approx 0.33$ produces a metal-to-insulator transition, whereas our
analysis of a quantum site percolation model predicts $y_c \approx 0.83$.
One hypothesis that is consistent
with these results is that since strong correlations are ignored in our 
quantum site percolation model, which includes the effects of configurational 
disorder only, such strong electronic correlations are both present and 
important.
\end{abstract}


\maketitle

The oxide spinel \LiTiO~has been the subject of considerable
experimental and theoretical study.
It was first synthesised and structurally characterised in 1971 by 
Deschanvres {\em et al.}\cite{Deschanvres71} Superconductivity, 
at 11K, was identified in 1973 by one of the present authors and his
collaborators.\cite{Johnston73} A comprehensive study of the normal state
and superconducting properties of \LixTiO~
(for $0 \leq x \leq 1/3$) was reported in 1976,\cite{Johnston76I,Johnston76II}
and a superconducting transition temperature of 13K was observed.
A recent review\cite{Moshopoulou99} highlights many of the advances made since 
then.

There are several reasons to study this system. Firstly, it is interesting to 
note that superconductivity among spinel systems is very rare; {\textit e.g.}, 
of the 300 or so known spinels,\cite{Moshopoulou99} only four of them are 
superconductors - CuRh$_2$Se$_4$~(T$_c = 3.49$~K), CuV$_2$S$_4$~(T$_c = 4.45$~K), 
CuRh$_2$S$_4$~(T$_c = 4.8$~K), and \LiTiO~(T$_c = 11.3~K$) - and only one of 
these four is an oxide. So, that oxide, \LiTiO, has the highest transition 
temperature of any spinel. Secondly, conduction in this system is believed
to take place on the Ti sublattice via the $t_{2g}$ orbitals, as suggested,
\textit{e.g.}, by electronic structure 
calculations,\cite{Satpathy87,Massidda88} and these sites form a 
corner-sharing tetrahedral lattice(CSTL). Thus, this system represents an 
example of conduction on a \textit{fully frustrated} three-dimensional lattice.
Also, in this paper we will argue, supported by considerable experimental
evidence, that the conduction paths of \LixTiO~and \LiAlTiO~ are excellent 
physical realizations of 
\textit{quantum site percolation}.\cite{Kirkpatrick72,Shapir82}

Furthermore, and central to the motivation for our work, these same electronic 
structure calculations\cite{Satpathy87,Massidda88} point out that this is a 
narrow band electronic system, 
with the bandwidth of the $t_{2g}$ bands of the order of 2-3 eV, thus suggesting
that perhaps strong electronic correlations are present and important.
Indeed, others have reached similar conclusions;
notably, the phase diagram of Alex M\"uller,\cite{Muller96} summarising a view 
of how the increased strength of electronic correlations in transition metal 
oxides leads to higher and higher superconducting temperatures, includes the
Lithium Titanate system. Although the original experiments
and analysis suggested a weak-coupling BCS-like s-wave 
superconductor,\cite{Johnston76II} it was later suggested\cite{Heintz89} 
that off stoichiometry this material is in fact an ``anomalous" 
superconductor (although this claim is not without criticism\cite{Annett99}). 
We also mention that photoemission studies of
Edwards, \textit{et al.}\cite{Edwards84} are interpreted in terms of strong
correlations, and magnetic susceptibility \cite{Harrison84} and specific heat 
data\cite{Heintz89} are interpreted in terms of a density of states that is 
moderately to strongly enhanced (see Ref.\cite{Dunstall94} for a 
discussion of these and other experiments). Taken together, these experimental 
results form a reasonably strong case for the presence of electronic 
correlations that are important to the physics of these materials.

Lastly, we mention the recent discovery of the first $d$-electron heavy fermion
compound, \LiVO.\cite{Kondo97} This system also assumes the spinel structure,
but so far no superconductivity has been observed. The active transition metal 
ion in \LiVO~has a formal valence of $d^{1.5}$, whereas for \LiTiO~one 
considers $d^{0.5}$ ions. Thus, \LiTiO~is a lower electronic density system than
\LiVO, and an understanding of its behaviour would seem to be a prerequisite 
to a full understanding of the Vanadate material. For example, why does
\LiTiO~superconduct, with a relatively high T$_c$, while \LiVO~does not
superconduct at all?

In attempt to gain more understanding of the \LiTiO~system, and, in
particular, to try and understand whether or not strong electronic correlations
are present, we have examined the density driven metal-to-insulator transition 
(MIT) of the related \LixTiO~and \LiAlTiO~compounds; for $x_{MIT}\sim 0.12$
and $y_{MIT} \sim 1/3$, 
transitions\cite{Johnston76I,Harrison85,Heintz89,Lambert87,Lambert90} 
to a non-metallic state (which we refer to as insulating) are encountered.
To be specific, we use a one electron 
approach to study this transition employing a quantum site percolation model.
Our work may be viewed as addressing the question of whether or not the 
MIT undergone by this system is driven by disorder only, similar to an 
Anderson-like MIT.  We find that the answer is no, and thus this work provides
indirect theoretical support for the proposal that strong electronic 
correlations are important in a description of the complicated transitions 
undergone in the \LiTiO~class of materials.

To fully explain our model we note the following: (i) Electronic
structure calculations\cite{Satpathy87,Massidda88} show that
the bands arising from the Ti $3d$ orbitals are separated from the 
O $2p$ band by about 2.4 $eV$; thus, the electronic valence state may be 
represented as Li$^{+1}$(Ti$^{+3.5}$)$_2$(O$^{-2}$)$_4$, and we ignore the 
oxygen sites and focus on only the Ti sites. The crystal octahedral field 
around Ti cations splits the Ti $3d$ bands into two separate and nonoverlapping
$t_{2g}$ and $e_g$ bands, with the $e_g$ bands split off above the $t_{2g}$ 
bands. Thus, formally this is a very low filling system --- 1/12th filling
of each of the (approximately) degenerate $t_{2g}$ bands. Although we have 
generalized our work to include all three $t_{2g}$ bands, here we will present
results for a one-band model of the Ti sublattice, and thus the stoichiometric 
compound is represented by a 1/4-filled band. (ii) Crystallographic refinements 
of the excess Li system \LixTiO~and the doped Al system \LiAlTiO~have consistently
demonstrated that both the excess Li and doped Al ions enters 
substitutionally onto the Ti sublattice (octahedral sites of the spinel 
structure).\cite{Deschanvres71,Johnston76I,HarrisonPhD,LambertPhD,Lambert90}
Assuming that the Li/Al ions that are substituted into the corner-sharing
tetrahedral lattice are fully ionised, these sites would block any
conduction electrons from hopping onto such sites; \textit{e.g.}, a simple 
argument supporting this follows from noting that the 
Li$^+$ ions will be at least doubly negatively charged relative to
the occupied Ti$^{3+}$ and unoccupied Ti$^{4+}$ sites that would exist in the
absence of substituting Li, and thus electrons will avoid these sites in 
favour of the Ti sites. We shall assume that these Li-substituting sites 
are removed from the sites available to the conduction electrons, which
implies that this system represents an excellent physical realization of site
percolation. 

Using such a model the simplest approach to characterising the MIT would be to 
then associate the transition with the critical concentration at which the 
classical percolation threshold is reached. For corner-sharing tetrahedral 
lattices, we have completed a large scale Hoshen-Kopelman\cite{Hoshen76} 
search, and have determined that this concentration corresponds to a 
probability of finding an occupied site at the transition of 
$p_c\sim 0.39 \pm 0.01$. Noting that the relationship between the
probability $p$ of site being occupied by a Ti ion (in the stoichiometric
Ti sublattice), and the excess Li concentration $x$, is given by
$x = 2(1-p)$, with an identical $y=2(1-p)$ relation for Al added to the Ti 
sublattice, in contrast to previous statements\cite{Harrison85,Lambert87} 
this $p_c$ corresponds to a critical excess Li, or added Al, of 
$x_c=y_c=2(1-p_c) \sim1.2$; that is Li$_{2.2}$Ti$_{0.8}$O$_4$ 
or LiAl$_{1.2}$Ti$_{0.8}$O$_4$. These very high levels of doping are well 
beyond the observed $x_{MIT} \sim 0.15$ or $y_{MIT} \sim 0.33$ concentrations 
at which the MITs occur. In fact, such a system would require large positive
Ti valencies well beyond anything seen in nature! Thus, the physics of 
these MITs is more complicated than simply the loss of an infinite maximally 
connected cluster at $p_c$.

We now consider a more accurate model for this system, a so-called quantum site
percolation model, which includes the dynamics of the electrons hopping on the
conducting, disordered sublattice of Ti sites. To be specific, 
the near-neighbour tight-binding Hamiltonian for \LiTiO~is
\begin{equation}
\label{eq:TB}
\hat{\mathcal{H}}= \sum_i \varepsilon_i c_i^\dagger c_i - t \sum_{<ij>} 
(c_i^\dagger c_j + h. c.)
\end{equation}
where $i$ labels the sites of the (ordered) Ti sublattice, 
$c_i^\dagger$ ($c_i$) is the creation (annihilation) operator of an electron 
at site $i$, $\sum_{<ij>}$ represents the sum over all nearest neighbour sites 
of a corner-sharing tetrahedral lattice, and $t$ is the near-neighbour hopping 
energy. To produce our model of quantum site percolation for the doped 
systems, the on-site energy $\varepsilon_j$ is determined by 
the probability of occupation, denoted by $p$, of a site being 
either a Ti ion, or a Li or Al dopant ion:
\begin{equation}
\label{eq:SiteDistr}
P(\varepsilon_i) = p\delta(\varepsilon_i-\varepsilon_{\rm Ti})+
(1-p)\delta(\varepsilon_i-\varepsilon_{\rm X}) \qquad
\end{equation}
where $\varepsilon_{\rm Ti}$ ($\varepsilon_{\rm X}$) is the on-site energy when
an electron occupies a Ti (X = Li or Al) site. In order to enforce that 
itinerant electrons move only on Ti sites we set $\varepsilon_{\rm Ti}=0$ and 
$\varepsilon_{\rm X}\to\infty$, and this limit connects this system with  
a quantum site percolation Hamiltonian.\cite{Kirkpatrick72,Shapir82}
Such considerations lead to the introduction of the quantum percolation 
threshold,\cite{Shapir82} usually denoted by $p_q$.
To be specific, $p_q$ is reached when all single-electron
energy eigenstates of the above tight-binding Hamiltonian are localized. 
That means that $p_q$ is always larger than $p_c$ since the presence of 
extended states necessarily requires an infinite maximally connected cluster.
Further, the evaluation of this quantity is warranted since $p_q > p_c$ implies
that the theoretical predictions of $x_c$ and $y_c$ given above would be
reduced by quantum percolation.

Our evaluation of $p_q$ for the corner-sharing tetrahedral lattice proceeds
as follows. We have considered various realizations of site percolated
lattices for several system sizes for a range of dopant concentrations;
to be specific, we consider lattices of size two, four, six, and eight cubed
conventional unit cells (noting that there are 16 Ti sites in the ordered
lattice per conventional unit cell), and then for each $p$ we examine
100, 50, 20 and 10 realizations consistent with this $p$, for two,
four, six, and eight cubed lattices, respectively.
For each realization we first apply the Hoshen-Kopelman algorithm to identify
the maximally connected cluster. Then, we diagonalize the one-electron
Hamiltonian describing the electron dynamics on this cluster.
Then, to determine the localized \textit{vs.} delocalized nature of the 
single-electron wave functions for the maximally connected cluster, we have 
used the scaling of the relative localization length as a function of system 
size, as described by Sigeti \textit{et al.}\cite{Sigeti91} This localization 
length, for a particular eigenstate, is defined by
\begin{equation}
\lambda~=~\sum_{ij}~|\psi_i|^2~|\psi_j|^2~ d(i,j)
\end{equation}
where $|\psi_i|^2$ is the probability amplitude for this eigenstate at site $i$,
and $d(i,j)$ is the Euclidean distance between lattice sites $i$ and $j$.
Then, the relative localization length is just the ratio of this quantity
to that for a state having a uniform probability amplitude throughout the
entire maximally connected cluster (we denote the latter by $\lambda_0$) 
--- this ratio thus provides a useful 
measure of the effective size, or localization, of a particular eigenstate 
relative to a Bloch state on the same maximally connected cluster.
The utility of this quantity (and we will describe its use for 
another problem below) is that if the quantity decreases as the system size is 
increased, that eigenstate corresponds to one that is spatially localized; 
the opposite behaviour is expected for delocalized states. We have
used this quantity as a means of identifying $p_q$.

As a test of this method, we note that for three-dimensional lattices,
reliable estimates exist only for the simple cubic lattice, and these
were obtained with a variety of different methods --- \textit{e.g.},
see the discussion in Ref.\cite{Soukoulis92} 
A value of $p_q = 0.44 \pm 0.02$ was identified,\cite{Soukoulis92}
and we have found that our method reproduces this number.

Using this method we find a value of $p_q$ for corner-sharing tetrahedral
lattices (with near-neighbour hopping only) of $p_q~=~0.52 \pm 0.02$, and
if we then associate this quantity with the concentration at which
the MIT occurs in doped \LiTiO, one finds $x_c=y_c=2(1-p_q)=0.96$, 
which correspond to Li$_{1.96}$Ti$_{1.04}$O$_4$ and  
LiAl$_{0.96}$Ti$_{1.04}$O$_4$. Again, these concentrations
are much higher than the experimentally measured values.

The reason for both of these failures is clear 
and has been suggested before\cite{Harrison85} 
--- as the concentration of 
doped cations is increased, the density of available $3d$ electrons is 
decreased. In fact, for $x_{BI}\equiv0.33$ and $y_{BI}\equiv1.0$ these 
materials become insulators simply because the bands of all allowed states are 
empty (we refer to such a state as a Band Insulator (BI)). So, if a one-electron 
description is going to correctly reproduce the MIT, we must account for the 
changing of the electronic density with cation dopant concentration. 
(This notwithstanding, we needed to determine what values of $x_c$ and $y_c$
were predicted by $p_q$ in case they were less than either $x_{BI}$,
or $y_{BI}$, respectively. Clearly, they are not.) 

\begin{figure}
\includegraphics[width=8cm]{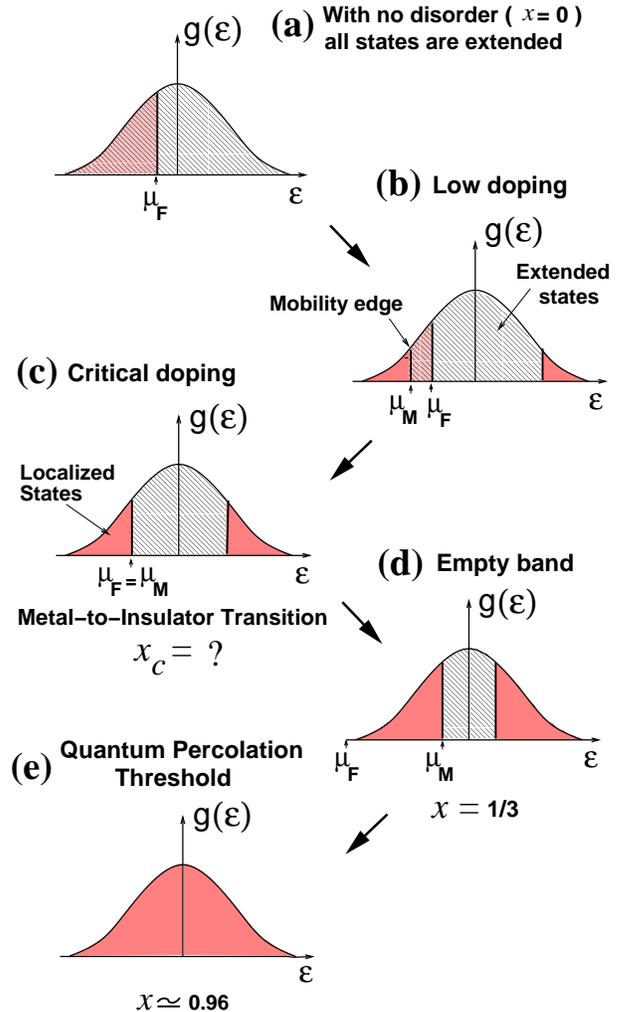}
\caption{\label{fig:MIT}A sequence of schematic diagrams that summarize 
how the MIT in 
\LixTiO would proceed {\em if} the transition was caused by disorder only.
The densities of states {\em vs.} energy are shown for several $x$. The
hatched regions identify the location of extended states, and the unhatched 
(solid) 
regions of density of states represent the location of localized states. (a) 
In \LiTiO all eigenstates are extended and the system 
is a $\frac14$-filled $d$ band conductor. (b) By doping Li cations randomly 
into the
Ti sites the system becomes disordered; thus, some energy eigenstates near 
the band edges
become localized, and, in particular, all states below the mobility 
edge $\mu_M$ are
localized. At the same time the density of itinerant electrons decreases, 
and thus the chemical potential $(\mu_F)$ is reduced. (c) When the chemical 
potential coincides with the mobility-edge, this model would predict that 
the MIT occurs. 
(d) The other end member of the homogeneity range of the spinel phase \LixTiO 
($0 \leq x \leq \frac13$) is an empty band insulator (note that the chemical 
potential
appears at the bottom of the allowed energy bands, and thus all electronic 
states are unoccupied). 
(e) The so-called quantum percolation threshold is reached when all electronic
states have a localized character (regardless of the location of the chemical 
potential).}
\end{figure}

Also, since the above arguments point to Ti occupation probabilities well above
the quantum percolation threshold, one is guaranteed to find a maximally 
connected cluster, and (possibly) several isolated clusters. By definition, 
all electronic states associated with an isolated cluster are localized, while 
for the maximally connected cluster both extended and localized states will be 
found. Similar to Anderson localization\cite{PWA58} of disordered systems, 
we expect that the eigenstates of the maximally connected cluster are localized 
in the band-edge and are separated from extended states at the middle of the band by 
the so-called mobility-edge.\cite{PWA58} By increasing disorder more and more
states become localized and the mobility-edge moves toward the centre 
of the band. However, as the concentration of cations (producing the disorder) 
is increased the density of itinerant $3d$ electrons is decreased. As long as 
the system's chemical potential (determined by both the density of $3d$ 
electrons and electronic density of states of the disordered system) is above 
the mobility edge, the system remains metallic, whereas if it was below the 
mobility edge, the states in the immediate vicinity of the Fermi surface would 
be localized and the system would display insulating behaviour. 
The critical dopant concentration at which the chemical 
potential and mobility edge meet thus identifies the MIT. The diagrams in 
Fig.~\ref{fig:MIT} summarize this process for the particular example of
\LixTiO (for which $x_c \sim 0.15$ and $x_{BI} \sim 0.33$).


\begin{figure}[t]
\includegraphics[width=8cm,height=7cm]{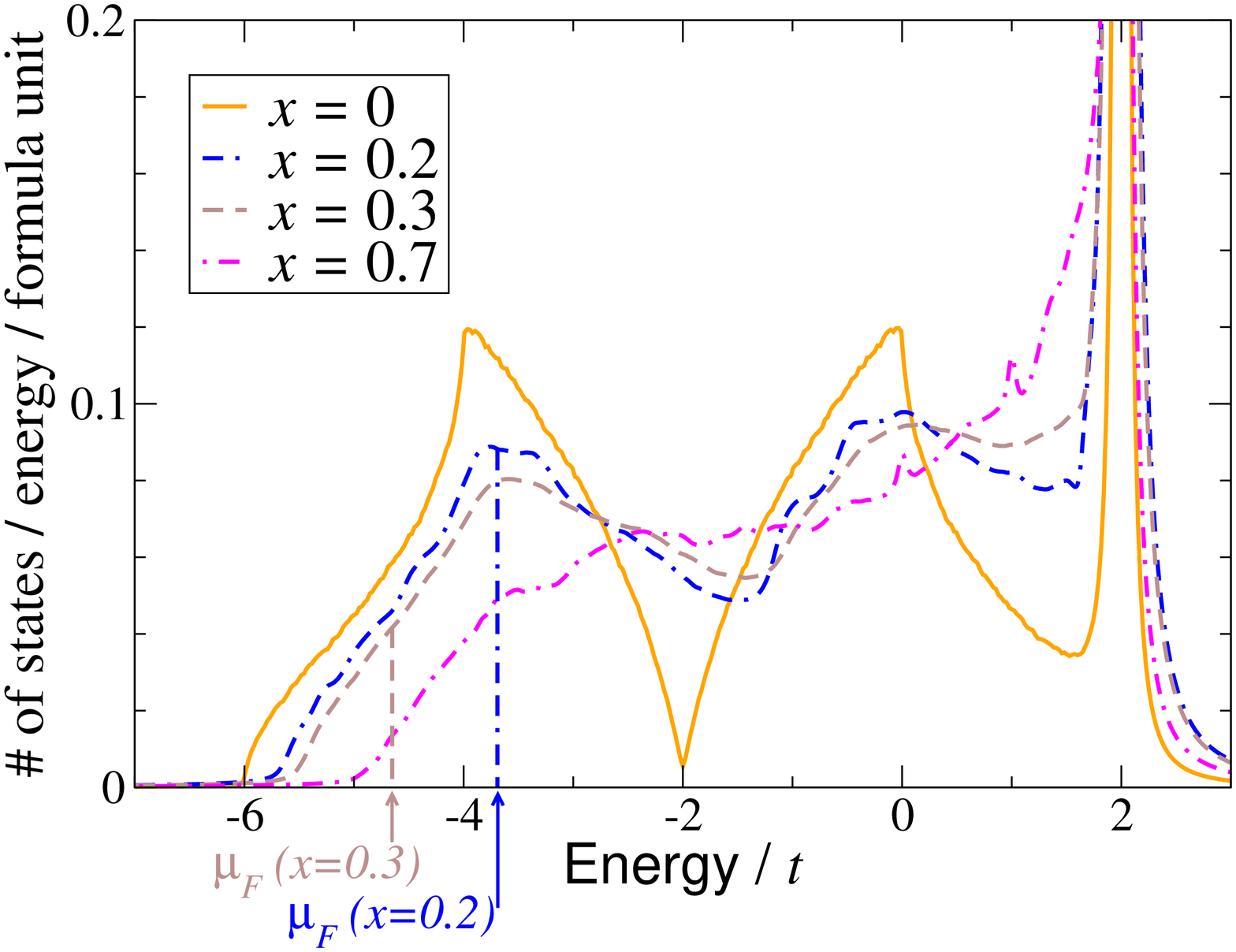}
\caption{\label{fig:DOS}Averaged numerical density of states for \LixTiO~
systems with 
different doping concentrations, and the corresponding location of chemical 
potentials. The DOS of the undoped system corresponds to the thermodynamic
limit, whereas plots for the doped systems are averaged density of states 
for 8$^3$ conventional unit cells using 10 different realizations of disorder.}
\end{figure}

Example density of states curves, and the locations of the chemical potentials,
are shown in Fig.~\ref{fig:DOS} for different doping concentrations,
averaged over systems with 8-cubed conventional unit cells (the largest 
studied).
Note that the full of spectrum of eigenvalues, of both the maximally 
connected cluster and all isolated clusters, have been evaluated since the 
latter contribute to the location of the chemical potential.

\begin{figure}[pt]
\includegraphics[width=8cm,height=7cm]{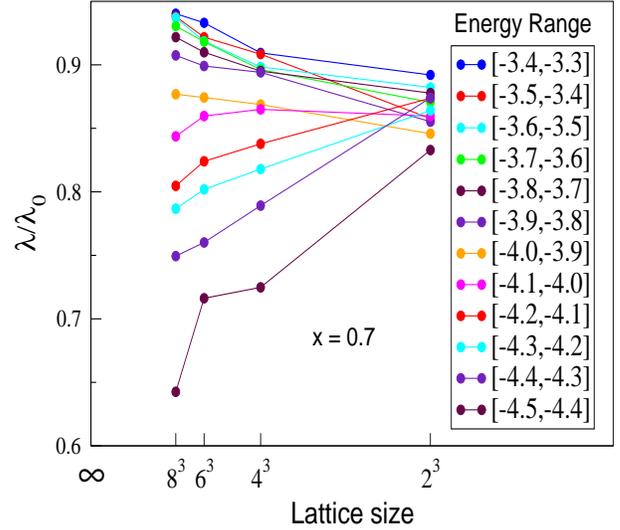}
\caption{\label{fig:scaling}Scaling of the relative localization length 
($\lambda$) of different energy ``bins" relative to that of the 
maximally connected cluster ($\lambda_0$), for energies close to the
mobility edge, for a doping concentration of $x = 0.7$. 
This ratio is plotted \textit{vs.} the reciprocal of the number of conventional 
cells, and as the size of the lattices is increased from 
2$^3$ conventional unit cells to 8$^3$ unit cells this
data shows that the eigenstates with energies above -3.9 have delocalized 
behaviour. For this system we estimate that the (lower) mobility edge
is located at $- 4.0\pm0.15$.}
\end{figure}

The mobility edge has been estimated using the above-mentioned scaling 
method,\cite{Sigeti91} and Fig.~\ref{fig:scaling} depicts the application of 
this method to identify the location of mobility edge for the specific doping 
concentration of $x = 0.7$. To be concrete, one can estimate that for
this dopant concentration the mobility edge, in units of the hopping energy 
$t$, is -4.0$\pm$0.15.

\begin{figure}[t]
\includegraphics[width=8cm,height=7cm]{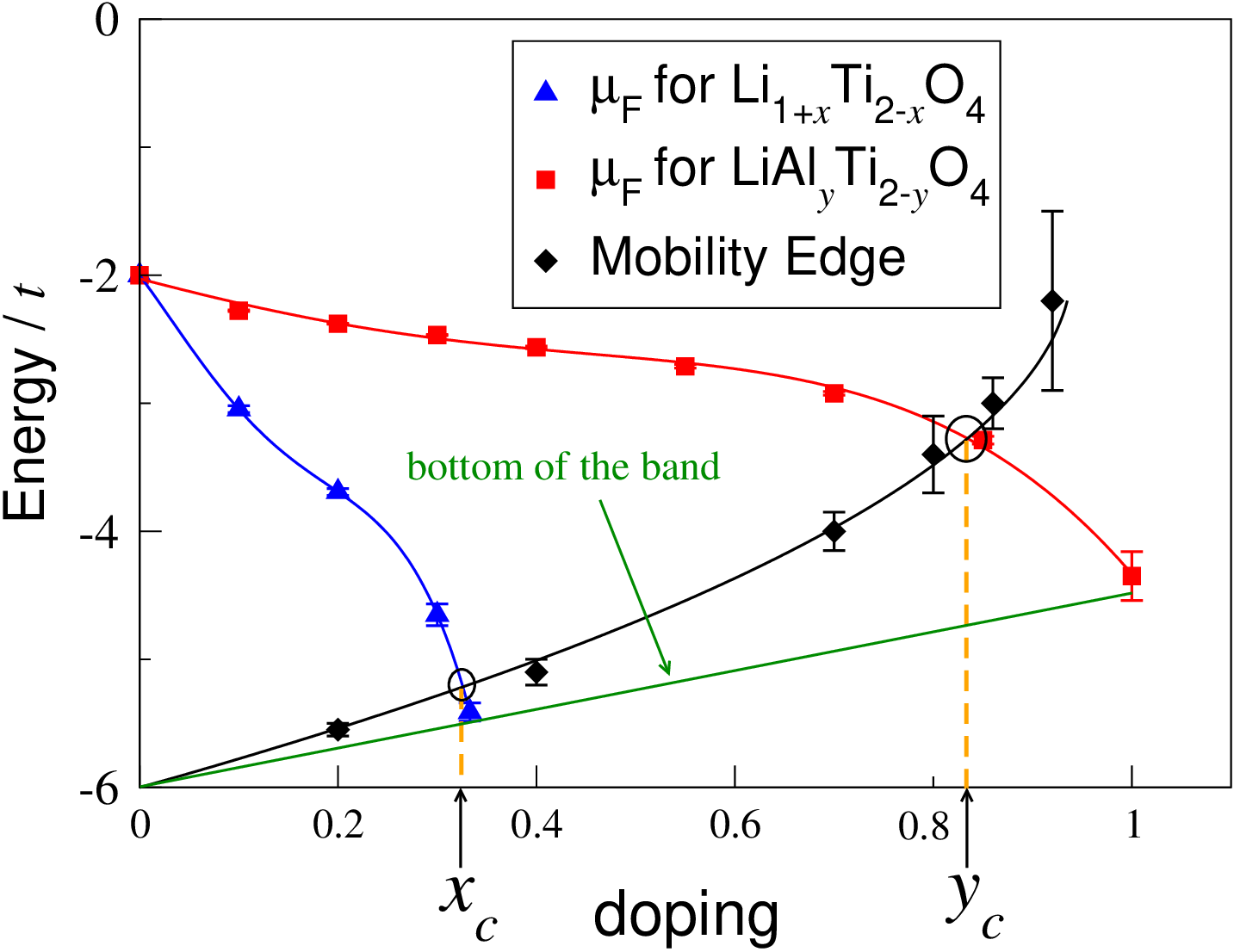}
\caption{\label{fig:bb_mu} This plot displays the final numerical results
of our study, and leads to the identification of predicted concentrations
at which the MITs occur. These data are the
the chemical potentials ($\mu_F$) and mobility edge as a 
function of doping for both \LixTiO~and \LiAlTiO; for reference we have
also plotted the minimum of the band of electronic states.
The crossing of the chemical potential and mobility edge (denoted by open black circles) would indicate the position of disorder-only 
induced metal-to-insulator transition, and these values are labelled
$x_c$ and $y_c$.}
\end{figure}

We have combined all of our data in Fig.~\ref{fig:bb_mu}, which shows the
chemical potential and mobility edge that we would estimate for both the
excess Li and doped Al systems. For reference we have included the values
of the band minimum for all dopings. Note that the Fermi level of the excess 
Li system crosses the mobility edge at roughly $x_c \approx 0.324$, whereas for
the doped Al system this crossing occurs at $y_c \sim 0.83$. Clearly,
these numbers are much larger than the experimental values ($x_{MIT} \approx 0.15$ 
and $y_{MIT} \approx 0.33$). Thus, we have studied a system that should be
very well represented, in a one-electron theory, by a quantum site
percolation model, have determined the concentrations at which the Fermi
levels cross the mobility edge, and find these values to be a factors of
2 and greater than 3 larger than the experimental results. So, we
believe that this shows that a one-electron model, that ignores electronic
correlations, cannot explain the observed MITs.

\newpage
Concluding, our results show that disorder-only models of the MITs undergone 
by these systems substantially overestimate the critical concentrations of 
doped cations. To be specific, for \LixTiO~our numerical result for
$x_c$ is a little more than double the experimental value, and for
\LiAlTiO~our numerical result is an even larger multiple. (We note that we have
not eliminated the possibility that polaronic effects give rise to this transition,
although no experimental work points to their existence.\cite{Moshopoulou99})
Thus, indirectly, this study supports the hypothesis that strong electronic 
correlations are important for the MIT, and, possibly, also for the 
$T_c \sim 13 K$ superconducting transition undergone by \LiTiO.

This past year\cite{chen03} sample preparation advances have allowed for the growth 
of large stoichiometric \LiTiO~ single crystals, and a 
remarkably sharp superconducting transition ($\delta$T$_c$ $\sim$ 0.1~K) 
has been observed. We hope that these new samples stimulate 
further experimental research in this interesting class of materials.

We wish to thank Gene Golub for several helpful comments. Part of
this paper was written when one of the authors (RJG) was visiting the Fields Institute
for Research in Mathematical Sciences, and he thanks them for their support
and hospitality. This work was supported in part by the NSERC of Canada, OGSST, 
and the USDOE. 

\bibliography{./LiTiO_MIT_2lanl}

\end{document}